\documentclass{article}
\usepackage{spconf,booktabs}
\usepackage{url}
\usepackage{breakurl} 
\usepackage[inline]{enumitem}

\urldef{\technical}\url{https://www.iter.org/album/Media/7%20-%20Technical}
\urldef{\woodaxes}\url{https://sound-effects.bbcrewind.co.uk/search?q=07054072%20OR%2007058026}
\title{Data leakage in cross-modal retrieval training: A case study}
\name{Benno Weck$^{1,2}$,
      Xavier Serra$^{2}$
      }
\address{$^1$ Huawei Technologies, Munich Research Center, Germany\\
        benno.weck@huawei.com\\
        $^2$ Universitat Pompeu Fabra, Music Technology Group, Spain\\
        benno.weck01@estudiant.upf.edu, xavier.serra@upf.edu\\
        }
\begin{document}
%
\maketitle
\begin{abstract}
The recent progress in text-based audio retrieval was largely propelled by the release of suitable datasets.
Since the manual creation of such datasets is a laborious task, obtaining data from online resources can be a cheap solution to create large-scale datasets.
We study the recently proposed SoundDesc benchmark dataset, which was automatically sourced from the BBC Sound Effects web page.
In our analysis, we find that SoundDesc contains several duplicates that cause leakage of training data to the evaluation data.
This data leakage ultimately leads to overly optimistic retrieval performance estimates in previous benchmarks.
We propose new training, validation, and testing splits for the dataset that we make available online.
To avoid weak contamination of the test data, we pool audio files that share similar recording setups. 
In our experiments, we find that the new splits serve as a more challenging benchmark.
\end{abstract}
\section{Introduction}
\label{sec:intro}
Retrieving audio through textual search queries was traditionally approached by extracting metadata from all audio files in the collection and selecting items by text-based matching algorithms.
With the advent of deep-learning-based methods, it became feasible to map search queries directly into the audio content domain in order to retrieve items at a large scale.
This form of content-based audio retrieval is commonly referred to as \textit{text-based audio retrieval}.

The research in this relatively young field is mostly driven by the availability of large-scale datasets.
These datasets serve as a source for the necessary training data and, additionally, allow for a comparative evaluation of different approaches.
For text--based audio retrieval, the most commonly used datasets are \textit{Clotho} \cite{drossos_clotho_2020} and \textit{AudioCaps} \cite{kim_audiocaps_2019}.
Both datasets were originally designed for the task of automatic audio captioning but lend themselves well to cross-modal retrieval. 
Nevertheless, there are certain drawbacks to both: 
\begin{enumerate*}[label=(\roman*), itemjoin={{, }}, itemjoin*={{ and }}]
  \item Clotho is limited in size and variability of the audio content and
  \item the audio content in AudioCaps is not freely accessible
\end{enumerate*}
The recently presented \textit{SoundDesc} dataset \cite{koepke_audio_2022} addresses both shortcomings: it is large in size while keeping a wide variation in content topics and its audio is freely available for research purposes. 
Moreover, the authors propose it as a benchmark for text-based audio retrieval.
The dataset was sourced from the BBC Sound Effects Archive website\footnote{\url{https://sound-effects.bbcrewind.co.uk/}}.
Due to the semi-automatic nature of the dataset creation process, it is more likely to contain undesired artefacts. 
We study the data distribution among the publicly available splits of the dataset and identify a couple of defects.
Primarily, we detect several duplicate recordings.
These duplicate recordings cause an unwanted overlap of the training and the evaluation part of the dataset, a so-called data leakage.
This leakage ultimately leads to overly optimistic retrieval scores reported on the test set.
We are convinced that these defects need to be corrected so that no erroneous conclusions are made and to maximise the potential of the data.
This is why, in this work, we set out to propose an updated version of the training, validation, and test splits of the SoundDesc dataset.
More specifically, our contributions are as follows:

\begin{itemize}
\itemsep=0em
\item We identify duplicates and overlapping recordings in the dataset using an off-the-shelf audio fingerprinting system.
\item We show that these duplicates lead to a data leakage problem and overly optimistic retrieval scores.
\item We propose new dataset splits that avoid weak contamination between development and evaluation data by grouping recordings that potentially share the same recording process.
\item We make the new splits available online.\footnote{\url{https://doi.org/10.5281/zenodo.7665917}}
\end{itemize}

\section{Related work}
\subsection{Text-based audio retrieval}

Text-based audio retrieval (sometimes also called language-based audio retrieval) can be described as the problem of ranking a set of audio files according to how closely they match a free-form query text.
This is a form of a cross-modal retrieval between two modalities, namely natural language text and audio.
For this task, text queries are usually provided as single-sentence descriptions of the audio, also referred to as captions.
A common approach in cross-modal retrieval is to employ a separate encoder model for each modality and map the respective outputs to a common representation space.
All submission in the Language-based audio retrieval task of the \textit{DCASE 2022} challenge \cite{xie_language-based_2022} used this model architecture.
Often these bi-encoder architectures rely on pretrained audio and text models \cite{koepke_audio_2022,weck_matching_2022, mei_metric_2022}.

\subsection{Dataset curation}
The goal of machine learning is to build a system that can generalise, i.e. perform well on previously unseen input data \cite{goodfellow_deep_2016}.
To judge this generalisation capability, machine learning practitioners usually keep a part of their data as a held-out set for evaluation.
This is commonly referred to as training and test splits of a dataset.
We usually assume that training and test data are independent of each other and identically distributed.
If information about the evaluation data is accessible to the machine learning model during training, these assumptions are violated and we speak of \textit{data leakage} \cite{kaufman_leakage_2012}.

This problem was studied in the context of audio datasets. 
For example, Sturm \cite{sturm_state_2014} uncovers several problems in a benchmark dataset widely used in music information retrieval research.
They find some exact duplicates in the dataset and show how certain characteristics of the data can be confounded with the ground-truth labels if left uncontrolled.
For instance, filtering the dataset so that all musical excerpts of an artist are kept in one split can have a significant impact on the performance measure.
This phenomenon is sometimes called ``artist effect'' \cite{flexer_effects_2010}.

As a positive example, Fonseca et al. explain their considerations when constructing the splits for the \textit{FSD50K} dataset \cite{fonseca_fsd50k_2022} -- a dataset collected from the online platform Freesound \cite{font_freesound_2013}.
They differentiate between data leakage and \textit{weak contamination}.
In their definition, this contamination can happen if items in a dataset are similar in some regard even if they are not the \textit{same}. 
Similar to what Sturm \cite{sturm_state_2014} describes about grouping songs of the same artist, they group recordings of the same content uploader.

\section{Data leakage in SoundDesc}
While employing the newly released SoundDesc dataset for our work on language-based audio retrieval, we accidentally stumbled across several duplicate audio items\footnote{For example: \woodaxes} in the proposed training and test splits.
This led us to performing a more thorough investigation about the general structure of this dataset and the proposed splits.

\subsection{The SoundDesc dataset}
The SoundDesc dataset was recently proposed by Koepke et al. \cite{koepke_audio_2022}.
It is a collection of audio recordings and sound effects sourced from the \textit{BBC Sound Effects Archive} website\footnote{\url{https://sound-effects.bbcrewind.co.uk/}} and contains 32979 audio files with associated textual descriptions.
Additionally, each item in the dataset is labelled with one primary category and potentially additional categories.
The authors of the dataset publish the training, validation and testing splits along with benchmark results on these splits.

\subsection{Detecting duplicates and overlapping recordings}
To automatically detect duplicates in the dataset, it requires a system to measure the similarity between recordings.
We employed the publicly available audio fingerprinting software \textit{Panako} \cite{six_panako_2022} since it was already successfully used to detect duplicates in a similar setting \cite{six_applications_2018}.
The \textit{Panako} algorithm was applied with its default configuration to generate a set of potentially matching audio recording pairs.
After manually reviewing a small subset, we decided to only keep pairs according to the following heuristic to filter out spurious matches:
Pairs are considered duplicates if
\begin{enumerate*}[label=(\roman*), itemjoin={{, }}, itemjoin*={{, and }}]
    \item the number of seconds containing sub-fingerprint matches exceeds 50\% of the total match duration
    \item a score of 25 or higher is required for a match to be considered.
\end{enumerate*}

We are left with 3601 distinct matching pairs.
We find that there are not only exact duplicates but also reprocessed recordings and recordings with partial overlaps.
These overlaps occur for example at the end and the start of another, or as one or more excerpts of a longer recording.
For simplicity, we will refer to all cases as \textit{duplicates}.

\subsection{The effect of duplicates on the benchmark results}
\label{ssec:results_duplicates}
After finding the duplicates in the dataset we want to assess their influence on the validity of the performance metrics measured in the published evaluation split.
To do so we first identify all pairs of duplicates that are split between the training and the evaluation part (validation \& testing) of the dataset.
We then create a new training set that is a subset of the original by excluding all items that have a duplicate recording in the evaluation data.
This way we can keep the test set unchanged and compare our results to the dataset creators' results.
In total, we exclude 1388 of the 23085 files in the training set.
To investigate the effect of the reduced training set size, we similarly construct three additional training sets by excluding the same number of files at random.
We refer to the new training subsets as \textit{deduplicated} and \textit{random}, respectively.

To assess the impact on the model training, in our experiments, we adopt the Collaborative-Experts (CE) \cite{liu_use_2019} model architecture as it was used by the SoundDesc authors in their benchmark \cite{koepke_audio_2022} experiment.
An implementation of this model is released together with the dataset.\footnote{\url{https://github.com/akoepke/audio-retrieval-benchmark/}}
After model training, we perform retrieval on the entire original test set.
Retrieval performance is measured as the recall at different levels by considering only the top $k$ elements for each query (R@$k$).
We report these metrics for the full test set and the subset of duplicates in the test set (648 items).
All results are given in Table \ref{tab:results} as the mean and the standard deviation computed over three runs with different random initialisations.

\begin{table}[th]
\centering
\begin{tabular}{@{}llll@{}}
\toprule
training set & R@1         & R@5         & R@10       \\ \midrule
             & \multicolumn{3}{c}{full test set}   \\ \cmidrule(l){2-4} 
original     & 31.3 $\pm$ 0.3   & 60.9 $\pm$ 0.6 & 70.9 $\pm$ 0.7     \\
deduplicated & 26.6 $\pm$ 0.6  & 55.5 $\pm$ 1.1  & 66.1 $\pm$ 0.5 \\\addlinespace
random 1     & 29.9 $\pm$ 0.5  & 59.1 $\pm$ 0.3  & 69.2 $\pm$ 0.3 \\
random 2     & 29.8 $\pm$ 0.4  & 58.5 $\pm$ 0.3  & 68.6 $\pm$ 0.5 \\
random 3     & 30.2 $\pm$ 0.1  & 59.2 $\pm$ 0.4  & 69.5 $\pm$ 0.1      \\ \addlinespace
             & \multicolumn{3}{c}{duplicates only} \\ \cmidrule(l){2-4} 
original     & 52.2 $\pm$ 0.9      & 82.8 $\pm$ 0.4      & 89.8 $\pm$ 0.6     \\
deduplicated      & 21.6 $\pm$ 0.6  & 49.7 $\pm$ 1.8  & 61.4 $\pm$ 0.9 \\\addlinespace
random 1     & 49.7 $\pm$ 2.5  & 80.0 $\pm$ 1.2  & 87.8 $\pm$ 0.6 \\
random 2     & 50.4 $\pm$ 0.6 &  81.5 $\pm$ 1.6 & 88.8 $\pm$ 0.5 \\
random 3     & 49.8 $\pm$ 1.4 & 81.0 $\pm$ 0.9 & 87.9 $\pm$ 1.2 \\ \bottomrule
\end{tabular}
\caption{Comparison of retrieval results for CE models trained with different training data}
\label{tab:results}
\end{table}

From the table, we can see that models trained without access to the duplicates in the training set (deduplicated) score significantly lower in all metrics than models trained with the full training set (original).
This large drop in performance is most likely only partially due to the fact that the former models have a smaller training set to learn from since models trained on a randomly reduced training set (random 1-3) only suffer a minor hit in performance.
This illustrates the influence of duplicates on the retrieval scores.
This effect is also apparent when only looking at the metrics measured on test items that have a duplicate in the training set (duplicates only).
Not surprisingly, we find that models can achieve significantly higher retrieval scores in the test subset of only duplicates than the entire test set when duplicates are left untreated (original).
For example, more than half of the duplicates are retrieved as the highest-ranked result (duplicates only/original R@1 = 52.2).
After deduplication, the scores reported for the subset (deduplicated) are even below the scores of the entire test set. 
These findings illustrate that there is a data leakage problem in the publicly available splits of the SoundDesc dataset that leads to overly optimistic benchmark results.
We argue that, in its current form, it can not be used as a benchmark dataset since it does not allow us to measure if any progress is made in solving the problem of text-based audio retrieval. 

\section{Proposing a new benchmark}
To establish a new benchmark on SoundDesc, the flaws discussed above need to be accounted for when splitting the data.
Additionally, we study if forms of weak contamination of the test data can be avoided using the metadata associated with the dataset.

\subsection{Treating duplicates}
As a minimal improvement, the discovered data leakage in SoundDesc needs to be fixed. Simply removing the duplicates from the training set would reduce the dataset size, which is not the preferred solution.
Instead, we propose to partition the data so that pairs of duplicates remain in the same split.
We form groups of recordings by assigning group labels to each pair of duplicates and by merging groups that share the same recording.
Items without any duplicates are left ungrouped.
Finally, we split the data and keep 15\% for validation and testing, each.
We use stratified sampling to maintain the same relative category distribution in all splits.
We refer to this split as \textit{clean}.

\subsection{Avoiding weak contamination}
A bulk of the SoundDesc is sourced from the BBC natural history unit (NHU) archive.
It comprises a large number of nature sounds, such as recordings of animals.
We surmise that some of these recordings might have a perceptual likeness even if they are no actual duplicates.
Especially recordings that share the same recording setting could have substantial similarities with each other.
For example, if multiple vocalisations of a bird are snippets of a long recording session, they might overlap in ambient noise, loudness, etc.
In the context of machine learning these overlaps can be considered unwanted artefacts since those artefacts could favour models that memorise rather than generalise well.
To avoid weak contamination of the test data in SoundDesc we propose to not split recordings with these kinds of potential overlaps when partitioning the dataset.

\begin{table*}[t]
\centering
\footnotesize
\begin{tabular*}{\linewidth}{@{}lllp{9.1cm}@{}}
\toprule
Date       & Recordist name & Topic (start of description)     & Description (rest)                                                                                  \\
\midrule
1996-11-21 & Graham Ross    & Camel Market                     & close-up mournful calls from camel. background voices in crowd.                                     \\
           &                & Camel Market                     & close-up calls from camel. Sounds of crowd \& individual voices.                                    \\
           &                & \ldots & \ldots                                                        \\ \addlinespace
           &                & Rajasthan Musicians              & medium close-up playing of flutes. Also sounds of drumming. Some background chatter from crowd.     \\
           &                & Rajasthan Musicians              & Sounds of masaks being played. Joined by singers \& later drums.                                    \\
           &                & \ldots & \ldots                                                        \\ \addlinespace

1977-05-31 & Lyndon Bird    & Green Tree Frog (Hyla Cinerea)   & Chorus close-up with crickets and distant traffic                                                   \\
           &                & Green Tree Frog (Hyla Cinerea)   & Chorus close-up with crickets                                                                       \\
           &                & \ldots & \ldots                                                                                                                        \\ \addlinespace
           & Roy Horton     & Common Bee Fly (Bombylius Major) & close-up hum from fly hovering. Birds in background.                                                \\
           &                & Common Bee Fly (Bombylius Major) & close-up hum from fly hovering. Birds in background. Also sheep, cockerel and voices in background. \\
           &                & \ldots                           & \ldots                                                                                              \\
\bottomrule
\end{tabular*}
\caption{Examples of metadata associated  with grouped recordings in our proposed split.}
\label{tab:group_examples}
\end{table*}

To automatically identify potential groups of recordings, we rely on metadata of the NHU archive:\footnote{This metadata is also accessible through the BBC~Sound~Effects~Archive~website but was not included in the SoundDesc dataset.} the date of the recording, the name of the recordist(s) and the topic of the recording.
The topic is given for all NHU recordings at the start of the description text and separated from the rest with a single dash. 
Recordings that do not have a recording date or a recordist name associated will be left ungrouped.
We can assign 12444 recordings to groups.
Table \ref{tab:group_examples} shows examples of the resulting groups.
We merge the newly defined groups with the groups of the \textit{clean} split and use the same stratification process to divide the data.
We refer to the resulting splits as \textit{group-filtered}.

\subsection{Benchmark results}
To set a benchmark for our proposed splits, we compare the previously introduced CE model with our own model.
Our system relies on a bi-encoder architecture that has shown promising results for the task of text-based audio retrieval \cite{weck_matching_2022}.
We follow the same training procedure described in our previous work \cite{weck_matching_2022} and only make few minor adjustments to the model architecture: 
As an audio encoder, we employ a pre-trained \textit{PANNs} model \cite{kong_panns_2020} that we keep fixed during training.
The audio embedding sequence extracted by this encoder is reduced by computing the mean and maximum across the time dimension and stacking the resulting vectors.
The stacked vectors are then mapped to a dimensionality of 768 using a multi-layer perceptron (MLP).
As text encoder, we employ a pre-trained \textit{distilroberta-base} \cite{liu_roberta_2019,sanh_distilbert_2019} model.
We take the first vector in the extracted text embedding sequence as the encoder result and similarly process it with an MLP.

\begin{table}[th]
\centering
\begin{tabular}{@{}llll@{}}
\toprule
training set & R@1           & R@5          & R@10         \\ \midrule
             & \multicolumn{3}{c}{clean split}          \\ \cmidrule(l){2-4} 
CE           & 27.3 $\pm$ 0.6    & 55.9 $\pm$ 0.4   & 66.5 $\pm$ 0.5   \\
Ours         & 28.0 $\pm$ 0.6    & 55.5 $\pm$ 0.7   & 65.6 $\pm$ 0.6   \\ \addlinespace
             & \multicolumn{3}{c}{group-filtered split} \\ \cmidrule(l){2-4} 
CE           & 20.4 $\pm$ 0.6    & 43.9 $\pm$ 0.2   & 53.7 $\pm$ 0.5   \\
Ours         & 20.8 $\pm$ 0.3    & 44.3 $\pm$ 0.8   & 54.2 $\pm$ 0.6   \\
\bottomrule
\end{tabular}
\caption{Comparison of retrieval results achieved by two different methods on our newly proposed SoundDesc splits. }
\label{tab:results_splits}
\end{table}

We use the same evaluation metrics as described above.
Table \ref{tab:results_splits} compares the results obtained by the two methods on each of our proposed splits.
It is apparent from the figures in the table that there are no significant differences between the results achieved by the different models in either of the splits.
As expected, the results for the clean split are in the range of the results of the experiments with the deduplicated training data discussed in Section \ref{ssec:results_duplicates}.
Interestingly, the retrieval scores are significantly lower when the group-filtered split is used.
This suggests that it is harder to retrieve the correct recording in this split of the data. 
A possible explanation for these results may be that there was indeed weak contamination of the test data present in SoundDesc and models could make use of overlaps in the data to solve the retrieval problem.

\section{Conclusion}
In this paper, we demonstrated that a data leakage problem in the publicly available splits of \textit{SoundDesc} leads to overly optimistic retrieval results.
Using an off-the-shelf audio fingerprinting software, we identified that the data leakage stems from duplicates in the dataset.
We define two new splits for the dataset: a \textit{cleaned} split to remove the leakage and a \textit{group-filtered} to avoid other kinds of weak contamination of the test data. 
From the results achieved by two different retrieval models, we conclude that our splits of the dataset serve as a more challenging benchmark for text-based audio retrieval.

\bibliographystyle{IEEEbib}
\bibliography{main}

\begin{thebibliography}{10}

\bibitem{drossos_clotho_2020}
Konstantinos Drossos, Samuel Lipping, and Tuomas Virtanen,
\newblock ``Clotho: an {Audio} {Captioning} {Dataset},''
\newblock in {\em {ICASSP} 2020 - 2020 {IEEE} {International} {Conference} on
  {Acoustics}, {Speech} and {Signal} {Processing} ({ICASSP})}, Barcelona,
  Spain, May 2020, pp. 736--740, IEEE.

\bibitem{kim_audiocaps_2019}
Chris~Dongjoo Kim, Byeongchang Kim, Hyunmin Lee, and Gunhee Kim,
\newblock ``{AudioCaps}: {Generating} {Captions} for {Audios} in {The}
  {Wild},''
\newblock in {\em {NAACL}-{HLT}}, 2019.

\bibitem{koepke_audio_2022}
A.~Sophia Koepke, Andreea-Maria Oncescu, Joao Henriques, Zeynep Akata, and
  Samuel Albanie,
\newblock ``Audio retrieval with natural language queries: A benchmark study,''
\newblock {\em IEEE Transactions on Multimedia}, pp. 1--1, 2022.

\bibitem{xie_language-based_2022}
Huang Xie, Samuel Lipping, and Tuomas Virtanen,
\newblock ``Language-based {Audio} {Retrieval} {Task} in {DCASE} 2022
  {Challenge},''
\newblock Accepted at DCASE 2022 Workshop, 2022.

\bibitem{weck_matching_2022}
Benno Weck, Miguel~Pérez Fernández, Holger Kirchhoff, and Xavier Serra,
\newblock ``Matching {Text} and {Audio} {Embeddings}: {Exploring}
  {Transfer}-{Learning} {Strategies} for {Language}-{Based} {Audio}
  {Retrieval},''
\newblock in {\em Proceedings of the 7th {Workshop} on {Detection} and
  {Classification} of {Acoustic} {Scenes} and {Events} 2022, {DCASE} 2022,
  {Nancy}, {France}, {November} 3-4, 2022}, Mathieu Lagrange, Annamaria
  Mesaros, Thomas Pellegrini, Gaël Richard, Romain Serizel, and Dan Stowell,
  Eds. 2022, Tampere University.

\bibitem{mei_metric_2022}
Xinhao Mei, Xubo Liu, Jianyuan Sun, Mark Plumbley, and Wenwu Wang,
\newblock ``On {Metric} {Learning} for {Audio}-{Text} {Cross}-{Modal}
  {Retrieval},''
\newblock in {\em Interspeech 2022}. Sept. 2022, pp. 4142--4146, ISCA.

\bibitem{goodfellow_deep_2016}
Ian Goodfellow, Yoshua Bengio, and Aaron Courville,
\newblock {\em Deep {Learning}},
\newblock MIT Press, 2016.

\bibitem{kaufman_leakage_2012}
Shachar Kaufman, Saharon Rosset, Claudia Perlich, and Ori Stitelman,
\newblock ``Leakage in data mining: {Formulation}, detection, and avoidance,''
\newblock {\em ACM Trans. Knowl. Discov. Data}, vol. 6, no. 4, pp. 15:1--15:21,
  2012.

\bibitem{sturm_state_2014}
Bob~L. Sturm,
\newblock ``The {State} of the {Art} {Ten} {Years} {After} a {State} of the
  {Art}: {Future} {Research} in {Music} {Information} {Retrieval},''
\newblock {\em Journal of New Music Research}, vol. 43, no. 2, pp. 147--172,
  2014.

\bibitem{flexer_effects_2010}
Arthur Flexer and Dominik Schnitzer,
\newblock ``Effects of {Album} and {Artist} {Filters} in {Audio} {Similarity}
  {Computed} for {Very} {Large} {Music} {Databases},''
\newblock {\em Comput. Music. J.}, vol. 34, no. 3, pp. 20--28, 2010.

\bibitem{fonseca_fsd50k_2022}
Eduardo Fonseca, Xavier Favory, Jordi Pons, Frederic Font, and Xavier Serra,
\newblock ``{FSD50K}: an open dataset of human-labeled sound events,''
\newblock {\em IEEE/ACM Transactions on Audio, Speech, and Language
  Processing}, vol. 30, pp. 829--852, 2022,
\newblock Publisher: IEEE.

\bibitem{font_freesound_2013}
Frederic Font, Gerard Roma, and Xavier Serra,
\newblock ``Freesound {Technical} {Demo},''
\newblock in {\em Proceedings of the 21st {ACM} {International} {Conference} on
  {Multimedia}}, New York, NY, USA, 2013, {MM} '13, pp. 411--412, Association
  for Computing Machinery,
\newblock event-place: Barcelona, Spain.

\bibitem{six_panako_2022}
Joren Six,
\newblock ``Panako: a scalable audio search system,''
\newblock {\em Journal of Open Source Software}, vol. 7, no. 78, pp. 4554, Oct.
  2022.

\bibitem{six_applications_2018}
Joren Six, Federica Bressan, and Marc Leman,
\newblock ``Applications of {Duplicate} {Detection} in {Music} {Archives}:
  {From} {Metadata} {Comparison} to {Storage} {Optimisation},''
\newblock in {\em Digital {Libraries} and {Multimedia} {Archives}}, Giuseppe
  Serra and Carlo Tasso, Eds., Cham, 2018, Communications in {Computer} and
  {Information} {Science}, pp. 101--113, Springer International Publishing.

\bibitem{liu_use_2019}
Y.~Liu, S.~Albanie, A.~Nagrani, and A.~Zisserman,
\newblock ``Use {What} {You} {Have}: {Video} retrieval using representations
  from collaborative experts,''
\newblock in {\em {arXiv} preprint arxiv:1907.13487}, 2019.

\bibitem{kong_panns_2020}
Qiuqiang Kong, Yin Cao, Turab Iqbal, Yuxuan Wang, Wenwu Wang, and Mark~D.
  Plumbley,
\newblock ``{PANNs}: {Large}-{Scale} {Pretrained} {Audio} {Neural} {Networks}
  for {Audio} {Pattern} {Recognition},''
\newblock {\em IEEE/ACM Transactions on Audio, Speech, and Language
  Processing}, vol. 28, pp. 2880--2894, 2020.

\bibitem{liu_roberta_2019}
Yinhan Liu, Myle Ott, Naman Goyal, Jingfei Du, Mandar Joshi, Danqi Chen, Omer
  Levy, Mike Lewis, Luke Zettlemoyer, and Veselin Stoyanov,
\newblock ``{RoBERTa}: {A} {Robustly} {Optimized} {BERT} {Pretraining}
  {Approach},'' 2019.

\bibitem{sanh_distilbert_2019}
Victor Sanh, Lysandre Debut, Julien Chaumond, and Thomas Wolf,
\newblock ``{DistilBERT}, a distilled version of {BERT}: smaller, faster,
  cheaper and lighter,'' 2019.

\end{thebibliography}

\end{document}